\documentclass[twocolumn,showpacs, amssymb, amsmath, nobibnotes, aps, pre]{revtex4-1}
\usepackage{graphicx}
\usepackage{dcolumn}
\usepackage{bm}
\usepackage{epsfig}
\usepackage{amssymb}
\usepackage{psfrag}
\usepackage{subfigure}
\usepackage{color}
\usepackage{longtable}
\usepackage[T1]{fontenc}
\usepackage{natbib}

\usepackage{array}
\newcolumntype{x}[1]{>{\centering\hspace{0pt}}p{#1}}%

\begin{document}
\title{Three dimensionality in quasi-two dimensional flows: recirculations and \emph{Barrel effects}}

\author{A. Poth\'erat}%

\affiliation{
Applied Mathematics Research Centre, Coventry University,
Priory Street, Coventry CV1 5FB, United Kingdom.}
\email{alban.potherat@grenoble-inp.org}

\date{\today}
%
\begin{abstract}
A scenario is put forward for the appearance of three-dimensionality both in quasi-2D rotating flows and quasi-2D  magnetohydrodynamic (MHD) flows. We show that 3D recirculating flows and currents originate in wall boundary layers and that, unlike in ordinary hydrodynamic flows, they cannot be ignited by confinement alone. They also induce a second form of three-dimensionality with quadratic variations of velocities and current across the channel. This scenario explains both the common tendency of these flows to two-dimensionality and the mechanisms of the recirculations through a single formal analogy covering a wide class of flow including rotating and MHD flows. These trans-disciplinary effects are thus active in atmospheres, oceans or the cooling blankets of nuclear fusion reactors.
\end{abstract}

\maketitle
%
\section{Introduction}
\begin{figure}
\psfrag{ex}{$\mathbf e_x$}
\psfrag{ey}{$\mathbf e_y$}
\psfrag{ez}{$\mathbf e_z$}
\psfrag{F}{$\mathbf F$}
\psfrag{z0}{$z=0$}
\psfrag{zm1}{$z=-1$}
\psfrag{H}{$H$}
\psfrag{eH}{$\epsilon_\nu H$}
\psfrag{uzh0}{$\check{u}_z(-1)$}
\includegraphics[width=8.5cm]{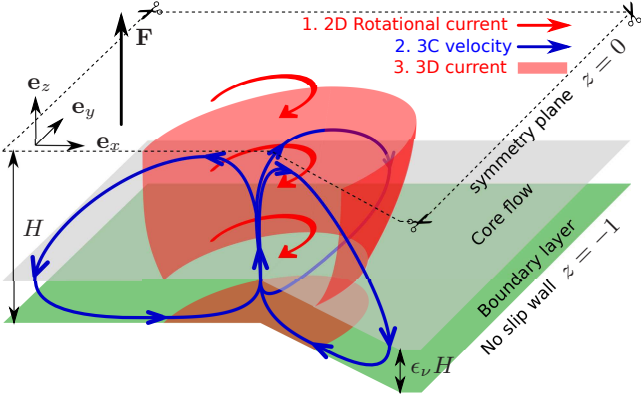}
\caption{Channel flow configuration. The three steps of the first Barrel 
effect are schematically represented. In the second type of Barrel 
effect, active in MHD flows, the steps are identical with the roles of the 
current and the velocity swapped.}
\label{fig:geo}
\end{figure}

Rapidly rotating flows and electrically conducting flows in homogeneous static  magnetic fields share a remarkable property: both tend to  two-dimensionality. In the former, inertial waves propagation promotes 
invariance along the rotation axis \cite{pedlosky87}. In the latter, when 
magnetic field perturbations induced by flow motion are negligible 
(in the \emph{quasi-static MagnetoHydroDynamic} (MHD) approximation), electric eddy 
currents 
damp velocity variations along the magnetic field lines \cite{roberts67}.\\
This feature is crucial 
because 2D and 3D flows have radically different 
dissipative and transport properties, especially when turbulent. 
While in 3D turbulence, energy follows a direct cascade from 
 large to small structures where it is dissipated by viscous friction, 
2D turbulence proceeds through an inverse cascade that accumulates energy in 
large structures where it is dissipated by friction on the boundaries 
\cite{tabeling02_pr,clerx09_amr}. In planar fluid layers, this  mechanism is 
suppressed as three-dimensionality emerges \cite{shats10_prl} and its 
suppression or appearance are respectively determined by the presence or 
absence of fluid motion along the third component 
\cite{celani10_prl, xia11_nphys}. The \emph{a priori} distinct questions of the 
number of components of velocity field (2C/3C) and the of spatial directions in which it varies are thus tightly linked and determine  whether flows obey 2D or 
3D dynamics, how and when they may switch between them.\\
The fundamental difference between 2D and 3D states places these questions at 
the centre a vast array of problems across disciplines: atmospheric and oceanic flows, electromagnetic flow control in metallurgical processes, even the dynamo problem.
Despite their importance, the transition mechanisms between 2D and 3D states, 
in rotating, MHD or other flows are not understood yet, and an important gap 
exists between theory and experiments.
Numerical simulations show that with periodic or free-slip boundaries, 
instabilities can break down strictly 2D MHD structures and lead to 
2D-3D intermittency \cite{thess07_jfm, boeck08_prl}. In rotating flows, 
non-linear transfer occurs from 3D to strictly 2D modes  
\cite{smith99_pf}. In experiments, shallow layers cannot be strictly 2D but 
only quasi-2D, 
either because of the three-dimensionality introduced by viscous boundary 
layers or because of variations of the same order in the core flow: firstly, 
such quasi-2D flows ignite 3D recirculations, that are crucial to the 2D-3D 
transition because of their influence on the inverse cascade and also because 
they nurture small-scale 3D turbulence \cite{sous04_pf}. 
 Secondly, in MHD channel flows, eddy currents 
between the boundary layer and the core were shown to induce a quadratic 
three-dimensionality \cite{muck00,psm00}, an effect recently observed 
experimentally, too \cite{kp10_prl}. Currently, the mechanisms that promote 
one or the other of these effects are unknown. 
Viscous boundary layers most certainly play a role in promoting 
3D recirculations in decaying and steady vortical flows \cite{satjin10_pf}, 
with and without background rotation \cite{akkermans08_pf}. Nevertheless, in 
the absence of Coriolis or Lorentz force, confinement alone suffices to trigger them \cite{akkermans08_epl}, even in the absence of boundary layer friction. These results stress the need for a clarification of the 
roles both of external forces such as the Lorentz or the Coriolis force on the 
one hand,  and  of confining boundaries on the other hand.\\ 
In this Letter, we examine the simple configuration of a symmetric, 
plane channel in a transverse field, bounded by two no-slip walls distant of 
$2H$ (fig. \ref{fig:geo}), to reveal how walls 
induce three-dimensionality in quasi-2D flows. The governing equations 
are written in a general form that defines a large class of flow 
encompassing not only quasi-static MHD and rotating flows, but also 
rotating-MHD flows relevant to geophysical dynamos. In doing so, a formal 
analogy is defined within this class of flows, which share a common tendency 
to two-dimensionality. 
A hierarchy of mechanisms is singled out that both 
explains how 3-Component (3C) motions are ignited and how three-dimensionality 
appears as a variation of physical quantities across the channel.\\
%
\section{Governing equations}

Following the formalism of rotating and quasi-static MHD flows, we first 
define a static homogeneous field $\mathbf F=F\mathbf e_z$ (vertical for convenience), orthogonal to the 
channel, that pervades the fluid domain and interacts with a current 
$\mathbf c$ to exert a force $\mathbf c\times \mathbf  F$ on the flow. In rotating flows ({\em resp.} quasi-static MHD), $\mathbf F$ is twice the background rotation vector ({\em resp.} the magnetic field) and $\mathbf c$ is the flow current ({\em resp.} electric current), that appear in the Coriolis ({\em resp.} Lorentz) force.
The current is linked to the velocity field $\mathbf u$ by a phenomenological 
law such as Ohm's law in MHD, that takes the general form $\mathbf c=\mathcal F(\mathbf u)$.
The corresponding non-dimensional equations for an incompressible fluid  (density $\rho$, 
viscosity $\nu$) depend on two dimensionless groups, built on $H$, the 
reference velocity $U$ and reference current $C$: $\epsilon=\rho U^2/(HCF)$ 
and $\epsilon_\nu^2=\rho\nu U/( H^2CF)$ respectively express the ratio of 
 inertia and viscous forces to $\mathbf c\times\mathbf F$, and both are smaller than unity. This class of analogy is summarised in table \ref{tab:anal}. We focus on regimes where deviations to the leading order play a key role, and 
in particular those due to inertia, \textit{i.e} typically 
$\epsilon\lesssim1$ and $10^{-2}\lesssim\epsilon_\nu<<1$.
Denoting any quantity $g$ in the core (\emph{i.e.} outside the wall boundary layers) as $\check g$, the equations write:
\begin{eqnarray}
\epsilon\frac{\check d}{dt}\check {\mathbf  u}_\perp+\mathbf \nabla_\perp \check p
&=&\check{\mathbf c}_\perp\times \mathbf e_z +\epsilon_\nu^2 (\partial^2_{zz}+\nabla_\perp^2) \check{\mathbf u}_\perp,
\label{eq:ns_perp}\\
\epsilon\frac{\check d}{dt} \check u_z+\partial_z \check p
&=&\epsilon_\nu^2 (\partial^2_{zz}+\nabla_\perp^2) \check u_z,
\label{eq:ns_z}\\
\partial_z \check{ u}_z&=&-\mathbf \nabla_\perp \cdot \check{ \mathbf u}_\perp,
\label{eq:cont_u}\\
\partial_z \check{ c}_z&=&-\mathbf \nabla_\perp \cdot \check{\mathbf  c}_\perp,
\label{eq:cont_c}\\
\check{\mathbf c}&=&{\mathcal F}(\check{\mathbf u}).
\label{eq:pheno}
\end{eqnarray}
Vector quantities and operators are split into their components along $\mathbf e_z$ (subscript $z$) and in the $(x,y)$ plane (subscript $\perp$). 
In rotating flows, $\check{\mathbf c}$ disappears from the governing equations, as (\ref{eq:pheno}) simply becomes $\check{\mathbf c}=\check{\mathbf u}$. In low-$Rm$ MHD, on the other hand, (\ref{eq:pheno}) represents Ohm's law:
\begin{equation}
\check{\mathbf c}=-\nabla\check\phi+\check{\mathbf u}\times\mathbf e_z,
\label{eq:ohm}
\end{equation}
where the core electric potential $\check\phi$ is usually obtained as solution of  the Poisson equation that follows from substituting (\ref{eq:ohm}) into 
(\ref{eq:cont_c}) \cite{muck00}.
%
\begin{table*}
\begin{tabular}{|x{1.5cm}|x{1.6cm}|x{2.9cm}|x{2.5cm}|x{2.5cm}|x{4.6cm}|}
\hline
		& Field $\mathbf F$	&current $\mathbf c=\mathcal F(\mathbf u)$&	$\epsilon$	& $\epsilon_\nu$& Wall boundary layer ($\hat{\mathbf u}^0_\perp=)$\tabularnewline

\hline
Quasi-static& magnetic field &electric current& Stuart number & Hartmann number & Hartmann layer \cite{moreau90}\tabularnewline

MHD&$ \mathbf B$& $\mathbf j=\sigma(-\nabla\phi+\mathbf u\times\mathbf B)$ &
$\quad N=\epsilon^{-1}=\frac{\sigma B^2H}{\rho U}$ & 
$\quad Ha=\epsilon_\nu^{-1}=BH\sqrt{\frac{\sigma}{\rho\nu}}$&
{$\check{\mathbf u}^0_\perp(-1)(1-e^{-\zeta})$}\tabularnewline

\hline
Rotating &double rotation& flow current &Rossby number & Ekman number & Ekman layer \cite{pedlosky87}\tabularnewline

flows	& $2\mathbf \Omega$& $\rho\mathbf u$&
$R\!o=\epsilon=\frac{U}{2H\Omega}$&
$E=\epsilon_\nu^2=\frac{\nu}{2H^2\Omega}$&
\parbox{4.5cm}{$\check{\mathbf u}^0_\perp(-1)(1-e^{-\frac{\zeta}{\sqrt{2}}}\cos\frac{\zeta}{\sqrt{2}})
+\mathbf e_z\times\check{\mathbf u}^0_\perp(-1) e^{-\frac{\zeta}{\sqrt{2}}}\sin\frac{\zeta}{\sqrt{2}}$}\tabularnewline
%
\hline
\end{tabular}
\caption{Analogy table giving the expressions of generic dimensional quantities $\mathbf F$, $\mathbf c$, and non-dimensional numbers $\epsilon_\nu$ and $\epsilon$ for quasi-static MHD and rotating flows. $\sigma$ and $\phi$ are the electric conductivity of the fluid and the electric potential.}
\label{tab:anal}
\end{table*}
%
\section{Tendency to two-dimensionality in the core flow}
We shall first clarify the interplay between the respective dynamics of 
$(\mathbf c_\perp, c_z)$ and $(\mathbf u_\perp, u_z)$ in the core. At the leading order $(\epsilon^0, \epsilon_\nu^0)$ when $\epsilon\rightarrow0, \epsilon_\nu\rightarrow0$, (\ref{eq:ns_perp}) and (\ref{eq:ns_z}) readily imply that $\check{ \mathbf c}_\perp$ and $\check p$ are independent of $z$ (\emph{i.e.} 2D). 
Denoting leading order  quantities with the superscript $g^0$,
\begin{eqnarray}
\check{\mathbf c}_\perp^0&=&\mathbf e_z\times\nabla_\perp \check p^0,
\label{eq:c_core_0}\\
\partial_z\check p^0&=&0.
\end{eqnarray}
The problem symmetry and (\ref{eq:cont_c}) further imply that the current is exclusively horizontal:
\begin{equation}
\check c_z^0=0.
\label{eq:cz0_core}
\end{equation}
 Two-dimensionality of the core velocity isn't ensured at this 
point, but depends on the nature of (\ref{eq:pheno}). For the time being, 
we shall assume that $z$-dependence doesn't explicitly appear in $\mathcal F$.
In this case, $\check{\mathbf u}_\perp^0$ is indeed 2D. When applied to 
rotating and quasi-static MHD flows, this result recovers the property 
that both are quasi-2D in the absence of inertial and viscous effects in the core. 
The $z-$dependence of $\check{\mathbf c}_\perp$ appears from (\ref{eq:ns_perp}) and (\ref{eq:ns_z}):
\begin{equation}
\partial_z\check{\mathbf c}_\perp=\left(\epsilon \frac{\check d}{dt} -\epsilon_\nu^2 \nabla^2\right)
\left[\nabla_\perp \check { u}_z-\partial_z\check {\mathbf u}_\perp \right]\times \mathbf e_z.
\label{eq:cp_core}
\end{equation}
Since the leading order core velocity is 2D, 
$\check{\mathbf c}_\perp$ must in fact be 2D at least up to 
$\mathcal O(\epsilon,\epsilon_\nu^2)$, so that
 using (\ref{eq:cont_c}) and the problem symmetry yields:
%
\begin{equation}
\check c_z=-z\check c_z(-1)+\mathcal O(\epsilon,\epsilon_\nu^2). 
\label{eq:cz_core}
\end{equation}
Similarly, the problem symmetry and (\ref{eq:cont_u}) imply that 
\begin{equation}
\check u_z=-z\check u_z(-1)+\mathcal O(\epsilon,\epsilon_\nu^2).
\label{eq:uz_core}
\end{equation}
From (\ref{eq:cp_core}) and (\ref{eq:cz_core}), the appearance of $z-$ 
variations in $\check{\mathbf c}_\perp$ and $\check{c}_z$ is 
 determined by the flow and the current injected into the core from the 
boundary layers that develop along the walls $\check u_z(-1)$ and $\check c_z(-1)$. 
Only then, can $z-$ variations in $\check{\mathbf u}_\perp$ appear 
through (\ref{eq:pheno}), at the same order as $\check{\mathbf c}_\perp$. Therefore, we now need to analyse the wall boundary layers.\\
%

\section{Recirculations driven in the Boundary layer}
In the vicinity of walls, horizontal viscous friction must 
balance $\mathbf c\times \mathbf F$ to achieve the no slip boundary condition, 
and this imposes the thickness of the resulting boundary layer to scale as $\epsilon_\nu$ (respectively the {\em Ekman} and {\em Hartmann} layers in rotating and MHD flows). The governing equations are accordingly rewritten 
in the boundary layer near the wall $z=-1$ using stretched variable 
$\zeta=\epsilon_\nu^{-1} (z+1)$, 
to reflect that viscous friction becomes $\mathcal O(1)$. Denoting any quantity $g$ in this region as $\hat g$, 
\begin{eqnarray}
\epsilon\frac{\hat d}{dt}\hat {\mathbf  u}_\perp+\mathbf \nabla_\perp \hat p
&=&\hat{\mathbf c}_\perp\times \mathbf e_z +(\partial^2_{\zeta\zeta}+\epsilon_\nu^2 \nabla_\perp^2) \hat{\mathbf u}_\perp
\label{eq:ns_perp_bl},\\
\epsilon\frac{\hat d}{dt} \hat u_z+\partial_z \hat p
&=&(\partial^2_{\zeta\zeta}+\epsilon_\nu^2 \nabla_\perp^2) \hat u_z
\label{eq:ns_z_bl},\\
\partial_\zeta \hat{ u}_z&=&-\epsilon_\nu \mathbf \nabla_\perp \cdot \hat{ \mathbf u}_\perp
\label{eq:cont_u_bl},\\
\partial_\zeta \hat{c}_z&=&-\epsilon_\nu\mathbf \nabla_\perp \cdot \hat{\mathbf  c}_\perp
\label{eq:cont_c_bl},\\
\hat{\mathbf c}&=&{\mathcal F}(\hat{\mathbf u}).
\label{eq:pheno_bl}
\end{eqnarray}
The no slip condition at the wall is written
\begin{equation}
\hat{\mathbf u}_\perp(-1)=0,
\label{eq:noslip}
\end{equation}
while the boundary conditions for $\hat{\mathbf c}$ and $\hat{ u}_z$ shall be 
left unspecified for the time being, thus allowing for possible wall-injection of current or mass into the fluid. 
Core and boundary layer variables must also satisfy a matching condition. Since 
they are not explicitly expanded in powers of $\epsilon,\epsilon_\nu$, the general matching condition for any quantity $g$ put forward by \cite{kaplun54} simplifies to \cite{psm02}:
%
\begin{equation}
\lim_{z\rightarrow -1}\check g(z+1)=\lim_{\zeta\rightarrow\infty}\hat g(\zeta).
\label{eq:match}
\end{equation}

%
$\check{c}_z(-1)$ is obtained by 
integrating $\nabla_\perp\times (\ref{eq:ns_perp_bl})$ using 
(\ref{eq:cont_c_bl}),  
%
%
(\ref{eq:match}), noting that 
$\partial_\zeta\check{\omega}_z(-1)=\epsilon_\nu\partial_z\check{\omega}_z(-1)$: 
\begin{equation}
\check{c}_z(-1)=c^W_z+\epsilon_\nu\partial_\zeta{\omega}_z^W +\mathcal O(\epsilon\epsilon_\nu,\epsilon_\nu^2),
\label{eq:cz0}
\end{equation}
%
where the superscript $W$ refers to values taken at the wall. The current injected at the wall $c^W_z$ would be determined 
by a boundary condition for the current, here unspecified.
Eqs. (\ref{eq:cz0}) and (\ref{eq:cz_core}) reveal the two main mechanisms that can 
feed eddy currents in the core: injection of current at the wall and 
rotational wall friction contribute respectively at the leading order, and at 
$\mathcal O(\epsilon_\nu)$. Importantly, (\ref{eq:cp_core}) implies that 
although these recirculations make 
$\check{\mathbf c}$ 3C, they do not directly affect the $z-$ dependence of $\check{\mathbf c}_\perp(z)$, which remains 2D up to $\mathcal O(\epsilon,\epsilon_\nu^2)$ at least.\\
Strikingly, if $c^W_z\neq0$, (\ref{eq:cz0}) contradicts (\ref{eq:cz0_core}): a 
linearly $z$-dependent vertical current in the core leads to diverging 
horizontal currents that induce a rotational force $\check{\mathbf c}_\perp\times\mathbf e_z$, which cannot be balanced by the sole leading order pressure 
gradient in (\ref{eq:c_core_0}). To resolve this paradox, 
viscous or inertial effects must exist in the core to oppose $\check{\mathbf c}_\perp\times\mathbf e_z$. 
These can therefore not be neglected and, in fact, prevent two-dimensionality at the leading order. Quasi-static MHD provides a well understood manifestation of 
this effect: MHD flows are often driven by injecting electric current through 
point-electrodes embedded in an otherwise electrically conducting 
wall \cite{sommeria88}. Above such an electrode develops a vortex of rotation 
axis $\mathbf e_z$, with a viscous core  that is 3D at the 
leading order. Injecting current at the wall therefore directly prevents 
quasi-two dimensionality, at least locally.
Having now singled out this 
important effect, we shall assume $c^W_z=0$, unless otherwise specified, for the reminder of this Letter 
and focus on higher order effects.\\
At order $\mathcal O(\epsilon_\nu)$, (\ref{eq:cz0}) generalises two classic  
properties of Ekman and Hartmann boundary layers: in rotating 
flows, friction in the Ekman layer gives rise to secondary flows in the core 
$\check u_z(-1)\simeq\frac{\sqrt2}2E^{1/2}\check\omega^0_z(-1)$, by \emph{Ekman pumping} 
\cite{pedlosky87}. 
In quasi-static MHD, $\partial_\zeta \hat{\mathbf u}_\perp^W=\check{\mathbf u}_\perp(-1)+\mathcal O(\epsilon_\nu,\epsilon)$, so (\ref{eq:cz0}) expresses that vorticity in the core drives an electric current $\check j_z(-1)\simeq Ha^{-1}\omega^0_z(-1)$ 
out of the Hartmann layers \cite{moreau90} (these two results are recovered using the leading order solutions of (\ref{eq:ns_perp_bl}-\ref{eq:pheno_bl}), 
$\hat{ \mathbf u}_\perp$, given in table \ref{tab:anal}).\\

We shall now turn our attention to the determination of $\check u_z(-1)$, which 
controls $\partial_z\check{\mathbf c}_\perp(z)$.
%
%
%
%
First, from (\ref{eq:ns_z_bl}) and (\ref{eq:cont_u_bl}), the pressure is constant across the boundary layer:
\begin{eqnarray}
\nabla_\perp \hat p(\zeta)&=&\nabla_\perp\check p(-1)+\mathcal O(\epsilon\epsilon_\nu,\epsilon_\nu^2). 
\label{eq:p_bl}
\end{eqnarray}
The horizontal pressure gradient in the boundary layer thus results from 
 the balance of forces in the core. (\ref{eq:p_bl}) and (\ref{eq:ns_perp_bl}) express that 
each of these forces alters the local balance between viscous 
forces and $\hat{\mathbf c}\times\mathbf e_z$ in the boundary layer, and thereby
drives horizontal jets. Should these jets be horizontally divergent, they 
 in turn induce a vertical flow from the boundary layer to the core. 
%
The equation for $\hat u_z$ thus follows from $\nabla_\perp\cdot$(\ref{eq:ns_perp_bl}) and (\ref{eq:cont_u_bl}), 
%
%
and $\check u_z(-1)$ is obtained by integration, using 
(\ref{eq:match}) and (\ref{eq:uz_core}):
\begin{eqnarray}
(1-&\epsilon_\nu&)\check{ u}_z(-1)=u^W_z+\epsilon_\nu u^C_z + \epsilon_\nu\epsilon u^I_z
+\mathcal O(\epsilon_\nu^3),
\label{eq:uz0}\\
%
%
u^C_z&=&\int_0^\infty\int_{\zeta}^\infty\int_{\zeta_2}^\infty \nabla_\perp\times\left[\hat{\mathbf c}_\perp-\check{\mathbf c}_\perp(-1)\right]\cdot\mathbf e_z d\zeta d\zeta_2 d\zeta_1, \nonumber \\
u^I_z &=&\int_0^\infty\int_{\zeta}^\infty\int_{\zeta_2}^\infty \nabla_\perp\cdot\left[
\frac{\check d}{dt}\check{\mathbf u}_\perp(-1) 
-\frac{\hat d}{dt}\hat{\mathbf u}_\perp 
\right] d\zeta d\zeta_2 d\zeta_1.
\nonumber 
\end{eqnarray}
Eq. (\ref{eq:uz0}) singles out three possible origins of secondary flows 
between core and boundary layer:
the flow directly injected at the wall $u_z^W$ doesn't interact with the 
boundary 
layer and is integrally transmitted to the core where it affects the flow at 
the leading order. The symmetric boundary conditions at the walls imply 
through (\ref{eq:uz_core})
 that it creates two symmetric recirculations there.
Should no flow be injected at the wall ($u_z^W=0$), secondary flows can result from a rotational current as 
$\check{u}_z(-1)=\epsilon_\nu u^C_z+\mathcal O(\epsilon\epsilon_\nu,\epsilon_\nu^2)$. Unlike wall-injected flow, the corresponding 
vertical flow builds up from horizontally divergent jets in the boundary layer 
but recirculates in the core in the same way. It leads to  secondary 
flows at order $\epsilon_\nu$ there, a scaling they inherit from the 
thickness of the boundary layer where they are created. Finally, if the core 
current is curl-free at the leading order, then $u^C_z=0$. Inertial effects 
of order $\epsilon$ then take over as the dominant mechanism that drives jets 
in the boundary layers, resulting in a secondary flow $\check{u}_z(-1)=\epsilon\epsilon_\nu u^I_z+\mathcal O(\epsilon_\nu^3)$. Importantly, no recirculation 
occurs if $u_z^W=u_z^C=u_z^I=0$, so in contrast to quasi-2D flows not subject 
to an external homogeneous field $\mathbf F$ \cite{akkermans08_epl}, they are 
not triggered by confinement alone.\\
%
%
%
%
In rotating flows, $\mathbf c=\mathbf u$ 
and $\mathbf \nabla \times \mathbf c_\perp=\omega_z\mathbf e_z\neq0$ in general  
(see table \ref{tab:anal}): the 
mechanism responsible for secondary flows is the Ekman pumping mentioned earlier when analysing 
eddy currents fed by $\check c_z(-1)$. 
Still, application of  (\ref{eq:uz0}) again recovers the well-known result that $\check{u}_z(-1)=\epsilon_\nu u^C_z+\mathcal O(\epsilon\epsilon_\nu,\epsilon_\nu^2)=\frac{\sqrt2}2E^{1/2}\check\omega_z(-1)+\mathcal(R\!oE^{1/2},E)$\cite{pedlosky87}.\\
In quasi-static MHD, by contrast,  
$u_z^C=0$ 
so in the inertialess theory of the Hartmann 
layer, no flow escapes to the core \cite{moreau90}. When inertia is taken into account though, (\ref{eq:uz0}) 
yields $\check{u}_z(-1)=\epsilon\epsilon_\nu u^I_z+\mathcal O(\epsilon\epsilon_\nu,\epsilon_\nu^3)=-(5/6) Ha^{-1}N^{-1} \nabla\cdot\left[\check{\mathbf  u}_\perp^0(-1)\cdot\nabla \check{\mathbf  u}_\perp^0(-1)\right]$ \cite{psm00}.\\
%

\section{The "Barrel" effects}
Having expressed the current and mass flow that feed into the core, we are now in position to return to the core flow equations analysed at the beginning of 
this Letter and determine the conditions of appearance of $z-$dependence in 
core quantities. First, since $\check{\mathbf c}_\perp$ appears at a higher order than $\check{\mathbf u}_\perp$ in (\ref{eq:ns_perp}-\ref{eq:cont_u}), and 
since we further assumed that $z-$ dependence didn't appear explicitly in 
$\mathcal F$,  $z$-dependence cannot appear at lower order in  $\check{\mathbf u}_\perp$ than in  $\check{\mathbf c}_\perp$.
Consequently, at the first order at which it appears, the second term in the expression of $\partial_z\check{\mathbf c}_\perp$ (\ref{eq:cp_core}) vanishes.
%
%
By virtue of the symmetric boundary conditions, and depending on 
(\ref{eq:uz0}), 
three-dimensionality may appear under either of the three forms:
\begin{eqnarray}
\check{\mathbf c}_\perp&=&\check{\mathbf c}_\perp(0)+\epsilon \frac{z^2}2\frac{d}{dt}\nabla_\perp u^W_z+\mathcal O(\epsilon\epsilon_\nu,\epsilon_\nu^2),\\
\check{\mathbf c}_\perp&=&\check{\mathbf c}_\perp(0)+\epsilon\epsilon_\nu \frac{z^2}2\frac{d}{dt}\nabla_\perp u^C_z+\mathcal O(\epsilon^2\epsilon_\nu,\epsilon_\nu^3),\\
\check{\mathbf c}_\perp&=&\check{\mathbf c}_\perp(0)+\epsilon^2\epsilon_\nu \frac{z^2}2\frac{d}{dt}\nabla_\perp u^I_z+\mathcal O(\epsilon_\nu^2).
\end{eqnarray}
Physically, these equations express that the $z$-linear vertical flow created 
in the core by the mass flow ejected from the boundary layers builds up a 
$z$-quadratic pressure in the core. A quadratic current must in turn be drawn in the core, for the force $\check{\mathbf c}_\perp\times \mathbf e_z$ to be able to
balance the corresponding quadratic component of the  horizontal pressure 
gradient.  
The $z$-dependence of $\check{\mathbf u}_\perp$ is determined by the nature of 
$\mathcal F$: in rotating flows for example, 
$\check{\mathbf c}_\perp=\mathcal F(\check{\mathbf u}_\perp)=\check{\mathbf u}_\perp$.
 Consequently, $\check{\mathbf u}_\perp$
 also depends quadratically on $z$, because $\check{\mathbf c}_\perp$ does.
Core structures are thus not columnar as usually assumed in quasi-2D flows 
but rather \emph{barrel}-shaped.\\ 
%
%
The Barrel effect has not yet been observed as such in rotating flows, but 
was  discovered in the MHD numerical simulations of \cite{muck00} and 
predicted theoretically by \cite{psm00}. We shall now show that the same 
general mechanism is at play in both cases. To this end, we must invoke a 
second analogy between MHD and rotating flows and release the assumption that 
$\mathcal F$ doesn't explicitly depend on $z$:
in MHD, the non-dimensional expression of $\mathcal F$ is given by Ohm's law 
${\mathbf c}={\mathbf j}=-\nabla \phi+\mathbf u\times\mathbf e_z$ (table \ref{tab:anal}). 
At the leading order, 
$z$-dependence is still absent in $\mathcal F$, 
so all core quantities remain 2D. Application of 
(\ref{eq:cz_core}) and (\ref{eq:cz0}) provide the expression of the vertical 
current there: $\check{ c}_z=Ha^{-1}z\omega_z^0+\mathcal O(Ha^{-2},N^{-1})$. 
According to Ohm's law, this  vertical electric current induces a quadratic 
component of the electric potential $\phi= (z^2/2) Ha^{-1} \check\omega_z^0+\mathcal O(Ha^{-2},N^{-1})$, which introduces an explicit $z$-dependence in $\mathcal F$ at this order. Since,  however, (\ref{eq:cp_core}) still implies that 
 $\check{\mathbf c}_\perp$ must be 2D at $\mathcal O(Ha^{-1})$, Ohm's law 
demands that $\check{\mathbf u}_\perp$ be quadratic at $\mathcal O(Ha^{-1})$. 
%
%
This mechanism is analogous to the Barrel effect identified previously, with $j_z$, $\phi$ and Ohm's law 
taking over the respective roles of $u_z$, $p$ and (\ref{eq:ns_z}). The first 
Barrel effect was driven by vertical \emph{flows} out of the boundary layer 
into the core, induced by 2D \emph{rotational currents}.  The MHD 
Barrel effect, by contrast,  is driven by vertical \emph{currents} out of the boundary layer, due to 2D \emph{vorticity}. This second analogy 
underlines that although one common formalism explains why rotating and MHD flows are either 2D, quasi-2D or 3C, the origins of their "Barrel" three-dimensionality are still formally analogous but involve a distinct analogy.\\

\section{Influence of the boundary conditions}
Up to this point, the analysis has been confined to a symmetric channel flow 
bounded by no-slip walls. In the quest for a flow closer to strict 
two-dimensionality, however, these conditions were often altered in a number of
 experimental and numerical studies of quasi-2D flows. 
We shall now examine how the driving mechanism for 3D recirculations and barrel effects are affected by these different boundary conditions in two main cases:
\subsection{No slip bottom wall and upper free surface}
The upper wall at $z=1$ is replaced with a non-deformable free surface 
(\emph{i.e.} it remains flat). This configuration was 
studied experimentally and numerically by \cite{akkermans08_pf}. The author 
showed in particular that free surface deformation didn't play an important 
role in the appearance of three-dimensionality, thus justifying the relevance 
of this boundary condition. Further assuming that no current escapes 
through the free surface, the boundary conditions at $z=1$ become:
\begin{equation}
\partial_z\mathbf u_\perp(1)=0 \qquad u_z(1)=0  \qquad
\label{eq:ufs_bc_u}
c_z(1)=0.
\end{equation}
In the channel with upper and lower no-slip walls, the problem symmetry implied 
$\partial_z\mathbf u_\perp(0)=0$, $u_z(0)=0$, 
and $c_z(0)=0$
, so the flow with a non-deformable free surface is simply that in the lower 
half of the channel found previously and the physical mechanisms remain the same.
\subsection{Upper and a lower non-deformable free surfaces}
This second, more ideal variant was investigated numerically with and without 
background rotation \cite{akkermans08_epl, akkermans08_pf}.
(\ref{eq:noslip}) is replaced with 
\begin{equation}
\partial_z {\mathbf u}_\perp(-1)=0.
\end{equation}
The direct consequence is that the viscous boundary layer at $z=-1$ is no 
longer present, so this boundary condition applies directly to the 
core flow (\emph{i.e.} $\check{\mathbf u}=\mathbf u$ and $\check{\mathbf c}=\mathbf c$). Again, the boundary condition at $z=-1$ for $u_z$ and $c_z$ can be left 
unspecified, to allow for possible injection of mass and current through the 
bottom free surface (The injection can be made symmetric as for the channel flow, by relaxing the condition $c_z(1)=0$ locally). 
This configuration is also symmetric with respect to the $z=0$ plane so 
equations  (\ref{eq:cz_core}) and (\ref{eq:uz_core}) for the core remain valid.
From (\ref{eq:cp_core}), injecting current at $z=-1$ therefore still leads to 
current recirculations and to three-dimensionality, both at the leading order, 
and locally prevents quasi-two dimensionality. Injecting mass also feeds 
recirculations at the leading order, but incurs a barrel effect at $\mathcal 
O(\epsilon)$, also from (\ref{eq:cp_core}). 
 If $u_z(-1)=c_z(-1)=0$, it follows from (\ref{eq:cz_core}) and 
(\ref{eq:uz_core}) that $c_z=u_z=0$: both vertical flows and 
vertical current become severed from their only source in (\ref{eq:cz_core}) 
and (\ref{eq:uz_core}), so no recirculation can occur, and the flow remains 2C. 
Since the barrel effects are driven by these recirculations, they are shut down 
too and the flow is indeed strictly 2D. Such conditions are ideal but they 
reveal 
the crucial role played by the boundary layers as sources of three-dimensionality in the presence of an external field $\mathbf F$. This result comes in  
stark contrast to similar configurations without external field where confinement 
alone incurs three-dimensionality \cite{akkermans08_epl}.\\
In practice, \cite{akkermans10_pre} proposed to reduce the influence of the 
boundary layer by resting the active fluid layer 
on another one, acting as buffer. The interface between the two fluids was 
assumed non-deformable but the boundary condition for the bottom boundary of the active layer wasn't exactly that of a free surface. Local viscous friction 
still existed and a viscous boundary layer developed there, albeit incurring  
less lower friction than if a solid wall were present. The authors noticed 
reduced but persistent three-dimensionality. Our analysis shows that in the 
presence of an external field, the same effect is present, as reducing the 
friction at the bottom boundary in (\ref{eq:cz0}) would immediately damp both 
the current recirculations through (\ref{eq:cz_core}) and the subsequent barrel 
effect. 
%
\section{Conclusion}
In this Letter, a quantitative analysis of the conditions of appearance of three-dimensionality in quasi-2D flows under the influence of an external field was 
conducted. It was shown that in the absence of any three-dimensional forces, 
three-dimensionality appears under the two forms of 3D recirculations and 
transverse variation of fluid quantities. We called the latter "Barrel effect" 
and proved that it was driven by the former.
Although a second order correction in the asymptotic sense, the Barrel effect 
becomes crucial at the transition between 2D and 3D regimes.
In turbulent flows, three-dimensionality is indeed driven by inertia and 
therefore appears when $\epsilon$ becomes closer to unity. This implies that 
three-dimensional instabilities that lead to a fully 3D state may not 
develop on a base flow made of columnar structures, as in strictly 2D flows, 
but rather on a flow with a quadratic profile that cannot be neglected 
and whose stability properties are currently unknown. 
Finally, this Letter stresses that in quasi-2D fluid layers pervaded by 
a uniform transverse field $\mathbf F$ but subject to no explicitly 3D force, 
3D recirculations and flow three-dimensionality are linked and occur 
exclusively because of the presence of boundary layers along the confining 
walls. This complements the findings of \cite{akkermans08_pf,akkermans08_epl}, 
who showed that in the same configuration, with and without background 
rotation, three-dimensionality could still be observed in the absence of 
boundary layer friction either if the forcing was itself 3D, or without 
background rotation, because of confinement alone.

%
\acknowledgments
Financial support of the Deutsche Forschungsgemeinschaft (grant PO1210/4-1) is gratefully acknowledged.\\

\bibliographystyle{plain}
\bibliography{p12_epl}

\begin{thebibliography}{10}

\bibitem{akkermans08_pf}
R.~A.~D. Akkermans, A.~R. Cieslik, L.~P.~J. Kamp, R.~R. Trieling, Clercx~H. J.,
  and Van Heijst G. J.~F. ~.
\newblock The three-dimensional structure of an electromagnetically generated
  dipolar vortex in a shallow fluid layer.
\newblock {\em Phys. Fluids}, 20:116601, 2008.

\bibitem{akkermans10_pre}
R.~A.~D. Akkermans, L.~P.~J. Kamp, H.~J.~H. Clercx, and G.~J.~F. van Heijst.
\newblock Three-dimensional flow in electromagnetically driven shallow
  two-layer fluids.
\newblock {\em Phys. Rev. E}, (2):026314, 2010.

\bibitem{akkermans08_epl}
R.A.D Akkermans, L.P.J Kamp, H.J.H. Clercx, and G.J.F. {van Heijst}.
\newblock Intrinsic three-dimensionality in electromagnetically driven shallow
  flows.
\newblock {\em Europhys. Lett.}, 83(2):24001, 2008.

\bibitem{boeck08_prl}
T.~Boeck, D.~Krasnov, and A.~Thess.
\newblock Large-scale intermittency of liquid-metal channel flow in a magnetic
  field.
\newblock {\em Phys. Rev. Lett.}, (101):244501, 2008.

\bibitem{celani10_prl}
A~Celani, S~Musacchio, and D~Vincenzi.
\newblock Turbulence in more than two and less than three dimensions.
\newblock {\em Phys. Rev. Lett.}, 104(18):184506, 2010.

\bibitem{clerx09_amr}
H.~J.~H Clercx and G~van Heijst.
\newblock Two-dimenisonal navier-stokes turbulence in bounded domains.
\newblock {\em Appl. Mech. Rev.}, 62:1--25, 2009.

\bibitem{kaplun54}
S.~Kaplun.
\newblock The role of coordinate systems in boundary layer theory.
\newblock {\em ZAMP}, V-9:111--135, 1954.

\bibitem{kp10_prl}
R.~Klein and A.~Poth\'erat.
\newblock Appearance of three-dimensionality in wall bouded mhd flows.
\newblock {\em Phys. Rev. Let.}, 104(3), 2010.

\bibitem{moreau90}
R.~Moreau.
\newblock {\em Magnetohydrodynamics}.
\newblock Kluwer, 1990.

\bibitem{muck00}
B.~M\"uck, C.~G\"unter, and L.~B\"uhler.
\newblock Buoyant three-dimensional mhd flows in rectangular ducts with
  internal obstacles.
\newblock {\em J. Fluid Mech.}, 418:265--295, 2000.

\bibitem{pedlosky87}
J.~Pedlosky.
\newblock {\em Geophysical Fluid Dynamics}.
\newblock Springer Verlag, 1987.

\bibitem{psm00}
A.~Poth\'erat, J.~Sommeria, and R.~Moreau.
\newblock An effective two-dimensional model for mhd flows with transverse
  magnetic field.
\newblock {\em J. Fluid Mech.}, 424:75--100, 2000.

\bibitem{psm02}
A.~Poth\'erat, J.~Sommeria, and R.~Moreau.
\newblock Effective boundary conditions for magnetohydrodynamic flows with thin
  hartmann layers.
\newblock {\em Phys. Fluids}, pages 403--410, 2002.

\bibitem{roberts67}
P.H. Roberts.
\newblock {\em Introduction to Magnetohydrodynamics}.
\newblock Longmans, 1967.

\bibitem{satjin10_pf}
M.~P. Satijn, A.~W. Cense, H~Verzicco, H.~J.~H Clercx, and G.~J.~F. van Heijst.
\newblock Three-dimensional structure and decay properties of vortices in
  shallow fluid layers.
\newblock {\em Phys. Fluids}, 13(7):1932--1945, 2001.

\bibitem{shats10_prl}
M.~Shats, D.~Byrne, and H.~Xia.
\newblock Turbulence decay rate as a measure of flow dimensionality.
\newblock {\em Phys. Rev. Lett.}, page 264501, 2010.

\bibitem{smith99_pf}
L.~M. Smith and F~Walefe.
\newblock Transfer of energy to two-dimensional large scales in forced,
  rotating three-dimensional turbulence.
\newblock {\em Phys. Fluids}, 11(6):1608--1622, 1999.

\bibitem{sommeria88}
J.~Sommeria.
\newblock Electrically driven vortices in a strong magnetic field.
\newblock {\em J. Fluid Mech.}, 189:553--569, 1988.

\bibitem{sous04_pf}
D~Sous, N~Bonneton, and J~Sommeria.
\newblock Turbulent vortex dipoles in a shallow water layer.
\newblock {\em Phys. Fluids}, 16(8):2886--2898, 2004.

\bibitem{tabeling02_pr}
P.~Tabeling.
\newblock Two-dimensional turbulence: a physicist approach.
\newblock {\em Phys. Rep.}, 362:1--62, 2002.

\bibitem{thess07_jfm}
A.~Thess and O.~Zikanov.
\newblock Transition from two-dimensional to three-dimensional
  magnetohydrodynamic turbulence.
\newblock {\em J. Fluid Mech.}, 579:383--412, 2007.

\bibitem{xia11_nphys}
H~Xia, D~Byrne, G~Falkovich, and M~Shats.
\newblock Upscale energy transfer in thick turbulent fluid layers.
\newblock {\em Nature Physics}, 7:321--324, 2011.

\end{thebibliography}
\end{document}